\documentclass[12pt]{article}
\usepackage{amsmath,amssymb}
\usepackage{graphicx}
\usepackage{color}
\begin{document}
\baselineskip=16pt
\begin{titlepage}
\begin{flushright}
{\small SU-HET-06-2013}\\
{\small IPMU13-0232}
\end{flushright}
\vspace*{1.2cm}

\begin{center}

{\Large\bf Planck scale boundary conditions in the standard model with singlet scalar dark matter}
\lineskip .75em
\vskip 1.5cm

\normalsize
{\large Naoyuki Haba}$^1$,
{\large Kunio Kaneta}$^2$ and 
{\large Ryo Takahashi}$^1$

\vspace{1cm}
$^1${\it Graduate School of Science and Engineering, Shimane University, 

Matsue, Shimane 690-8504, Japan}\\
$^2${\it Kavli IPMU (WPI), 
The University of Tokyo,\\ Kashiwa, Chiba 277-8568, Japan}\\

\vspace*{10mm}

{\bf Abstract}\\[5mm]
{\parbox{13cm}{\hspace{5mm}
We investigate Planck scale boundary conditions on the Higgs sector of the 
standard model with a gauge singlet scalar dark matter. We will find that vanishing self-coupling and Veltman condition at the Planck scale are realized 
with the 126 GeV Higgs mass and top pole mass, 172 GeV $\lesssim M_t\lesssim$ 
173.5 GeV, where a correct abundance of scalar dark matter is obtained with 
mass of 300 GeV $\lesssim m_S \lesssim$ 1 TeV. It means that the Higgs potential
 is flat at the Planck scale, and this situation can not be realized in the 
standard model with the top pole mass.
}}

\end{center}
\end{titlepage}

\section{Introduction}

The Higgs particle has just been discovered at the Large Hadron Collider (LHC) 
experiment~\cite{Chatrchyan:2013lba,CMS}. In addition, the results from the 
experiment are consistent with the standard model (SM), and an evidence of new 
physics such as supersymmetry (SUSY) is not obtained. Currently, the 
experimental results strongly constrain the presence of SUSY at low energy 
although the minimal supersymmetric standard model (MSSM) is an attractive 
candidate for new physics beyond the SM. Thus, a question, ``How large is new 
physics scale?'', might become important for the SM and new physics.
 One can consider several scenarios such as high scale supersymmetric models or 
a scenario without SUSY in which the SM is valid up to the Planck scale 
$M_{\rm pl}$, etc.

As an example of the later scenario, it was pointed out that 
imposing a constraint that the SM Higgs potential has two degenerate vacua, in 
which one of them is at the Planck scale, leads to the top mass $173\pm5~{\rm 
GeV}$ and the Higgs mass $135\pm9~{\rm GeV}$~\cite{Froggatt:1995rt}. More recent
 work~\cite{Shaposhnikov:2009pv} showed that an asymptotic safety scenario of 
gravity predicts $126~{\rm GeV}$ Higgs mass with a few GeV uncertainty. In 
these two scenarios, the boundary conditions (BCs) of the vanishing Higgs 
self-coupling ($\lambda(M_{\rm pl})=0$) and its $\beta$-function 
($\beta_\lambda(M_{\rm pl})=0$) are imposed at the Planck scale. In addition to 
these two BCs, the work~\cite{Holthausen:2011aa} also discussed the Veltman 
condition (Str$M^2(M_{\rm pl})$=0) and the vanishing anomalous dimension of the 
Higgs mass ($\gamma_{m_h}(M_{\rm pl})=0$) at the Planck scale. It was found that 
the four BCs yield a Higgs mass range of $127-142~{\rm GeV}$. Thus, combining 
these BCs can interestingly predict values of the Higgs and top masses in the SM
 close to the experimental ones but a slightly heavier Higgs mass and/or lighter
 top mass than experimental ones are generally predicted from these BCs as shown
 in~\cite{Degrassi:2012ry} (see 
also~\cite{Bezrukov:2012sa,Alekhin:2012py,Masina:2012tz,Hamada:2012bp,Jegerlehner:2013cta,Buttazzo:2013uya,Masina:2013wja} for the latest 
analyses). BCs of $\lambda(M_{\rm pl})=0$ and Str$M^2(M_{\rm pl})$=0 mean that there
 exists an approximately flat direction in the Higgs potential,\footnote{In 
order to confirm the existence of the flat direction, one should know the full 
ultraviolet (UV) completion. In the work, we assume an UV theory yielding the 
BCs for the flat direction at the Planck scale.} which might be adopted to the 
Higgs inflation~\cite{Bezrukov:2007ep,Bezrukov:2009db,Bezrukov:2010jz,Giudice:2010ka,Kamada:2010qe,Kamada:2012se,Lee:2013nv,Allison:2013uaa,Salvio:2013rja,Hamada:2013mya}. 
In addition, the quadratic (logarithmic) divergence for the Higgs mass disappear
 at the Planck scale under the Veltman condition (the vanishing anomalous 
dimension $\gamma_{m_h}(M_{\rm pl})=0$). In this work, we investigate the three BCs
 in a gauge singlet extension of the SM.

One important motivation for the gauge singlet extension of the SM is that the 
SM does not include a dark matter (DM). In the extension, the gauge singlet 
scalar can be DM when the scalar has odd parity under an additional $Z_2$ 
symmetry~\cite{Silveira:1985rk} (see 
also~\cite{Veltman:1989vw,McDonald:1993ex,Burgess:2000yq}). Once a gauge singlet
 scalar is added to the SM, an additional positive contribution from new scalar 
coupling appears in the $\beta$-function of the Higgs self-coupling and the 
Veltman condition (and the anomalous dimension of the Higgs 
mass).\footnote{See~\cite{Davoudiasl:2004be,HKT} for discussions of the vacuum 
stability and triviality in the SM with a gauge singlet real scalar. See 
also~\cite{Chpoi:2013wga} and references therein for implications of the LHC 
data to models with an extra singlet scalar, \cite{Iso:2009ss,Iso:2012jn} for 
the classically conformal $U(1)_{B-L}$ extended SM,~\cite{Holthausen:2013ota} for
 a model of electroweak and conformal symmetry breaking.} This means that the 
discussion of the three BCs at the Planck scale is modified from the SM. Since 
it actually seems difficult to reproduce 126 GeV Higgs mass and $173.07\pm1.24$ 
GeV top pole mass~\cite{Beringer:1900zz}~(see also~\cite{CDF:2013jga,CDF}) at 
the same time (i.e. a slightly heavier Higgs and/or a lighter top masses than 
the experimental center values are required) under the above three BCs at the 
Planck scale in the SM, it is interesting to investigate if the BCs could be 
realized with the center values of the Higgs and top masses in the singlet 
scalar DM extension of the SM. In this work, we take the following setup: (i) We
 consider a simple framework, in which only one gauge singlet real scalar is 
added to the SM. (ii) The gauge singlet scalar is DM. (iii) All scalar quartic 
couplings in the model can be perturbatively treated up to the Planck scale.

The paper is organized as follows: In Section 2, we investigate the three BCs at
 the Planck scale in the above framework. As a result, we will find 
that the vanishing self-coupling and Veltman condition at the Planck scale are 
realized with the 126 GeV Higgs mass and top pole mass, 171.8 GeV $\lesssim 
M_t\lesssim$ 173.5 GeV, where a correct abundance of scalar dark matter is 
obtained with mass of 300 GeV $\lesssim m_S \lesssim$ 1 TeV. It means that the 
Higgs potential is flat at the Planck scale, and this situation cannot be 
realized in the SM with the top pole mass. Section 3 is devoted to the summary.

\section{Boundary conditions at the Planck scale}

We consider the SM with a gauge singlet real scalar $S$, and investigate the 
values of scalar quartic couplings at the Planck scale by solving 
renormalization group equations (RGEs) in the model. The relevant Lagrangian of 
the model and the RGEs for the scalar quartic couplings are given by
 \begin{align}
  &{\cal L}        =       {\cal L}_{\rm SM} + {\cal L}_S, \\
  &{\cal L}_{\rm SM} \supset -\lambda\left(|H|^2-\frac{v^2}{2}\right)^2, 
                            \label{SM} \\ 
  &{\cal L}_S      =       -\frac{\bar{m}_S^2}{2}S^2-\frac{k}{2}|H|^2S^2
                            -\frac{\lambda_S}{4!}S^4+(\text{kinetic term}), 
                            \label{S}
 \end{align}
and
 \begin{eqnarray}
  (4\pi)^2\frac{dX}{dt}=\beta_{X}~~~(X=\lambda,k,\lambda_S), \label{R}
 \end{eqnarray}
with
 \begin{align}
  &\beta_\lambda
    =\left\{
      \begin{array}{ll}
       0 & \mbox{for }\mu<m_H \\
       24\lambda^2+12\lambda y^2-6y^4-3\lambda(g'{}^2+3g^2)+\frac{3}{8}\left[2g^4+(g'{}^2+g^2)^2\right] & \mbox{for }m_H\leq\mu<m_S\\
       24\lambda^2+12\lambda y^2-6y^4-3\lambda(g'{}^2+3g^2)+\frac{3}{8}\left[2g^4+(g'{}^2+g^2)^2\right]+\frac{k^2}{2} & \mbox{for }m_S\leq\mu
      \end{array}
     \right., \label{lam} \\
  &\beta_k 
    =\left\{
      \begin{array}{ll}
       0 & \mbox{for }\mu<m_S\\
       k\left[4k+12\lambda+6y^2-\frac{3}{2}(g'{}^2+3g^2)+\lambda_S\right] & \mbox{for }m_S\leq\mu
      \end{array}
     \right.,  \label{k} \\
  &\beta_{\lambda_S}
    =\left\{
      \begin{array}{ll}
       0 & \mbox{for }\mu<m_S\\
       3\lambda_S^2+12k^2 & \mbox{for }m_S\leq\mu
      \end{array}
     \right., \label{h}
 \end{align}
respectively, where we assume that the Higgs mass $m_H$ is smaller than DM mass 
$m_S$.\footnote{If $m_S<m_H$, $\beta_\lambda$ is zero for $\mu<m_H$ and is given 
by the third line of right-hand side of Eq.~(\ref{lam}) for $m_H\leq\mu$.} $H$ 
is the SM Higgs doublet, $v$ is the vacuum expectation value (VEV) of the Higgs 
as 246 GeV, $y$ is the top Yukawa coupling, $t$ is defined as $t\equiv\ln\mu$, 
and $\mu$ is a renormalization scale within the range of $M_Z\leq\mu\leq M_{\rm 
pl}$. We also impose an additional $Z_2$ symmetry on the model. Only the gauge 
singlet scalar has odd parity while all the SM fields have even parity under the
 symmetry. We give some comments on properties of the three scalar quartic 
couplings obeying Eqs.~(\ref{R})$\sim$(\ref{h}): 
\begin{itemize}
\item The right-hand side of Eq.~(\ref{k}) is proportional to $k$ itself. Thus, 
if we take a small value of $k(M_Z)$, where $M_Z$ is the $Z$ boson mass, a 
change of value in the running of $k(\mu)$ is also small and remained in a small
 value. As a result, the running of $\lambda$ closes to that of the SM. 

\item It is known as the vacuum instability that the value of $\lambda$ becomes 
negative before the Planck scale in the SM with the experimental center values 
of the Higgs and top masses. This is due to the negative contribution from the 
top Yukawa coupling to the $\beta$-function of $\lambda$ as in Eq.~(\ref{lam}). 
The minimum in the running of $\lambda$ is around $\mathcal{O}(10^{17})$ GeV. It 
is also shown from NNLO computations~\cite{Degrassi:2012ry} that $\lambda$ can 
remain positive up to the Planck scale when $127~{\rm 
GeV}\lesssim m_h\lesssim130~{\rm GeV}$ for $M_t=173.1\pm0.6~{\rm GeV}$ (or when 
$171.3~{\rm GeV}\lesssim M_t\lesssim171.7~{\rm GeV}$ for $m_h=126~{\rm GeV}$).

\item Once the gauge singlet scalar is added to the SM, the additional 
contribution of $k^2/2$ with the plus sign appears in the $\beta$-function of 
$\lambda$. This contribution can lift the running of $\lambda$, and thus, 
$\lambda$ can be around zero at the Planck scale. 

\item The position of the minimum in the running of $\lambda$ comes to lower 
energy scale than $\mathcal{O}(10^{17})$ GeV by adding the gauge singlet scalar 
because the contribution of $k^2/2$ in Eq.~(\ref{lam}) becomes large at a high 
energy scale compared to the electroweak (EW) scale.

\item The realization of the vanishing $\lambda$ around the Planck scale by 
adding the gauge singlet scalar means that $\lambda$ becomes negative before the
 Planck scale due to the above third and fourth properties of $\lambda$. Then, 
$\lambda$ returns to zero.

\item The running of $\lambda_S$ is an increasing function of $t$ (or $\mu$). 
There is not a direct contribution from $\lambda_S$ to the $\beta$-function of 
$\lambda$ but the running of $\lambda_S$ affects on that of $\lambda$ through 
the running of $k$.
\end{itemize}

We investigate the case that the gauge singlet scalar is DM with the three BCs 
($\lambda(M_{\rm pl})=0$, $\beta_\lambda(M_{\rm pl})=0$, and Str$M^2(M_{\rm pl})=0$) in 
this model. Since we impose the odd-parity on the singlet scalar under the 
additional $Z_2$ symmetry, the singlet can be a candidate for DM. Thus, 
$\Omega_Sh^2=0.119$ must be reproduced in the case, where $\Omega_S$ is the 
density parameter of the singlet and $h$ is the Hubble parameter. 

\subsection{Vanishing Higgs self-coupling: {\boldmath $\lambda(M_{\rm pl})=0$}}

First, we consider the BC that $\lambda$ is zero at the Planck scale 
$M_{\rm pl}=10^{18}$ GeV, $\lambda(M_{\rm pl})=0$. The BCs of the Higgs self-coupling
 and top Yukawa coupling at low energy are taken as
 \begin{eqnarray}
  \lambda(M_Z)=\frac{m_h^2}{2v^2}=0.131,~~~y(M_t)=\frac{\sqrt{2}m_t(M_t)}{v},
  \label{BC}
 \end{eqnarray}
for the RGEs, where $m_h=126$ GeV is taken, $M_t$ is the top pole mass as 
$173.07\pm1.24$ GeV, and $m_t$ is the $\overline{\rm MS}$ mass as 
$160_{-4}^{+5}$ GeV~\cite{Beringer:1900zz}.\footnote{We also take the following 
values as~\cite{Beringer:1900zz}, $\sin^2\theta_W(M_Z)=0.231$, $\alpha_{\rm 
em}^{-1}=128$, $\alpha_s(M_Z)=0.118$ for the parameters in the EW theory, where 
$\theta_W$ is the Weinberg angle, $\alpha_{\rm em}$ is the fine structure 
constant, and $\alpha_s$ is the strong coupling, respectively.} 

Let us solve the RGEs, Eqs.~(\ref{R})$\sim$(\ref{h}). Gray dots in 
Fig.~\ref{fig} (a) show the region satisfying $|\lambda(M_{\rm pl})|<10^{-2}$ and 
$\Omega_Sh^2=0.119$. In the figure, the horizontal axis is the gauge singlet DM 
mass defined by $m_S\equiv\sqrt{\bar{m}_S^2+kv^2/2}$ and the vertical axis is the
 top pole mass. The bounds of top mass $173.07\pm1.24~{\rm GeV}$ are also 
depicted by the horizontal dashed lines. Figure~\ref{fig}~(b) is a typical 
example of the runnings of the scalar quartic couplings satisfying the above 
conditions (and the Veltman condition discussed later). The horizontal and 
vertical axes are the renormalization scale and the values of scalar quartic 
couplings, respectively. Black, blue, and red curves indicate the runnings of 
$\lambda$, $k$, and $\lambda_S$, respectively. Initial conditions for the 
corresponding RGEs are $k(M_Z)=0.24$, $\lambda_S(M_Z)=0.34$, $M_t=173$ GeV, and 
$m_S=800$ GeV with Eq.~(\ref{BC}).

It can be seen from Fig.~\ref{fig} (a) that $|\lambda(M_{\rm pl})|<10^{-2}$ can be 
satisfied in a region of $85~{\rm GeV}\lesssim m_S\lesssim1.1\times10^3~{\rm 
GeV}$ with the corresponding top pole mass, $171.8~{\rm GeV}\lesssim 
M_t\lesssim173.8~{\rm GeV}$. Such a DM mass region will be checked by the future
 XENON100 experiment with 20 times sensitivity of the current 
data~\cite{Cline:2013gha}. One can also see that a larger top mass requires a 
larger DM mass in the region of $m_S\gtrsim 10^2~{\rm GeV}$. This is due to the 
following reason: A larger top mass needs a larger value of $k$ in order to 
realize the tiny value of $\lambda$ at the Planck scale. And a larger $k$ 
requires a larger DM mass to give the correct abundance in the range of 
$m_S\gtrsim 10^2~{\rm GeV}$ (e.g., see~\cite{HKT,Cline:2013gha}).

In order to realize the correct abundance of DM in $m_S\lesssim10^3$ GeV, 
$k(M_Z)\lesssim0.3$ is needed. Thus, $k(M_{\rm pl})$ is not also large 
($k(M_{\rm pl})<1$) for the realization of DM. Since we have also imposed the 
condition of $0<\lambda_S(M_{\rm pl})<1$ in the analyses, the model can be 
described by a perturbative theory up to the Planck scale. On the other hand, 
the value of $\lambda_S$ does not actually affect on the abundance of DM. Thus, 
the region described by the gray dots in Fig.~\ref{fig} (a) is not changed even
 with the condition of $\lambda_S(M_{\rm pl})>1$.
 
 One might suggest the Higgs inflation by the use of the region satisfying 
$\lambda(M_{\rm pl})=0$, which is included in gray dots of Fig.~\ref{fig} (a), but
 it is not possible.\footnote{If one considers $\lambda(\mu)>0$ and very small 
$\lambda(M_{\rm pl})$, one would have a successful Higgs inflation with a 
non-minimal coupling of the Higgs field to the Ricci curvature scalar.} Since 
$\lambda(\mu)<0$ ($10^8~{\rm GeV}\lesssim\mu<M_{\rm pl}$) and $\lambda(M_{\rm 
pl})=0$, there is a global minimum of the potential between the EW and Planck 
scales. If one identifies the Higgs with the inflaton, the inflaton rolls 
downslope to the global minimum not to the EW one. Thus, one must consider the 
other inflation models.

The smallness of $\lambda(M_{\rm pl})$ predicting close values of the Higgs and 
top masses to experimental ones at low energy motivates one to investigate the 
BC of $\lambda(M_{\rm pl})=0$ and/or the possibility of the Higgs inflation. On 
the other hand, the value of $\lambda_S$ does not strongly affect the SM (Higgs 
and top masses), DM sectors, and other cosmology compared to that of $\lambda$ 
(and $k$ which determines the abundance of DM). Thus, we focus only on the BC of
 $\lambda(M_{\rm pl})=0$ in this work. If there could be phenomenological and/or 
cosmological motivations to impose $\lambda_S(M_{\rm pl})=0$, the discussions of 
the realization of the BC might also be interesting.

\subsection{Veltman condition: Str{\boldmath $M^2(M_{\rm pl})=0$}}

The Veltman condition, which indicates a disappearance of  the quadratic divergence on the 1-loop radiative correction to the bare Higgs mass, is modified to
 \begin{eqnarray}
  \frac{\mbox{Str}M^2}{v^2}
   \equiv6\lambda+\frac{k}{2}+\frac{3}{4}g'{}^2+\frac{9}{4}g^2-6y^2=0,
  \label{V}
 \end{eqnarray}
where the term $k/2$ in Eq.~\eqref{V} is new contribution from $k|H|^2S^2/2$ 
interaction in the SM with gauge singlet scalar. In the SM, the value of 
the left-hand side of Eq.~(\ref{V}) without $k/2$ term at the Planck scale is 
$-0.291$ when one takes $m_h=126$ GeV and $M_t=173.07$ GeV.

\begin{figure}[t]
\hspace{4.5cm}(a)\hspace{8cm}(b)\vspace{-0.7cm}

\begin{center}
\includegraphics[scale=0.98]{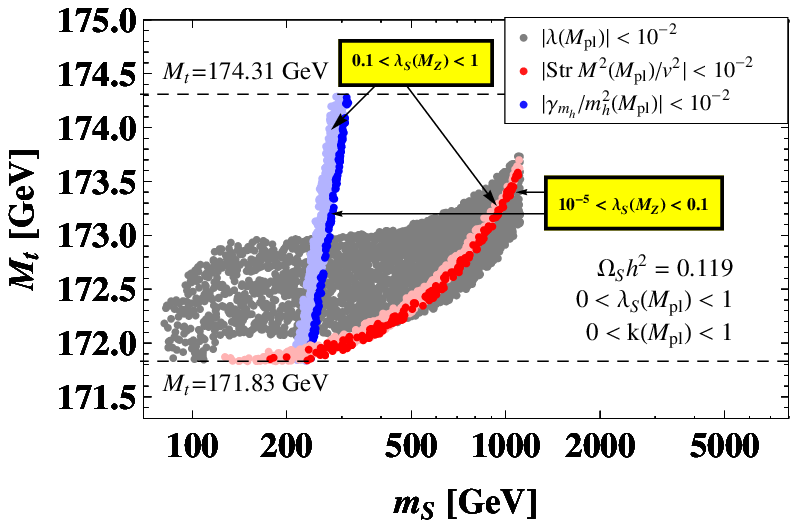}
\raisebox{3.5mm}{\includegraphics[scale=0.81]{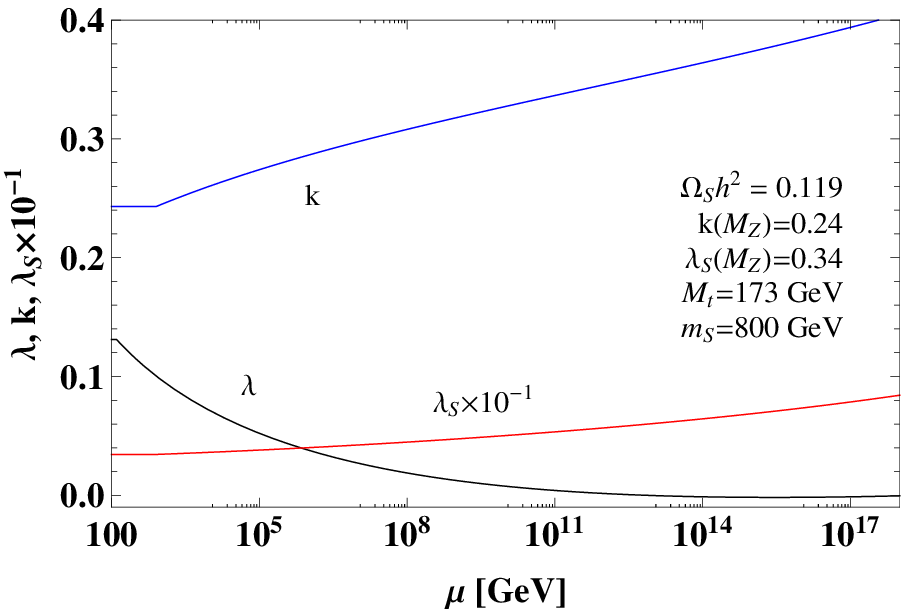}}
\end{center}
\caption{(a) Regions satisfying the BCs at the Planck scale, in which conditions
 $|\lambda(M_{\rm pl})|<10^{-2}$, $|{\rm Str}M^2(M_{\rm pl})/v^2|<10^{-2}$, and 
$|\gamma_{m_h}/m_h^2(M_{\rm pl})|<10^{-2}$ are depicted as gray, (deep and light) 
red, and (deep and light) blue, respectively. The deep and light red 
(blue) dots indicate $10^{-5}<\lambda_S(M_Z)<0.1$ and $0.1<\lambda_S(M_Z)<1$ for 
$|{\rm Str}M^2(M_{\rm pl})/v^2|<10^{-2}$ ($|\gamma_{m_h}/m_h^2(M_{\rm pl})|<10^{-2}$), 
respectively. (b) A typical example of runnings of $\lambda$, $k$, and 
$\lambda_S$, whose initial conditions are specified by $k(M_Z)=0.24$, 
$\lambda_S(M_Z)=0.34$. The parameter set of the figure (b) corresponds to a 
point of $(m_S,M_t)=(800~{\rm GeV}, 173~{\rm GeV})$ in the figure (a).
}
\label{fig}
\end{figure}

We also show the region satisfying $|\mbox{Str}M^2(M_{\rm pl})/v^2|<10^{-2}$ and 
$\Omega_Sh^2=0.119$ for the SM with the singlet DM in Fig.~\ref{fig} (a) by deep
 and light red dots. The deep and light red dots indicate 
$10^{-5}<\lambda_S(M_Z)<0.1$ and $0.1<\lambda_S(M_Z)<1$, respectively. One can see
 that $|{\rm Str}M^2(M_{\rm pl})/v^2|<10^{-2}$ can be satisfied in a region of 
$180~{\rm GeV}\lesssim m_S\lesssim1~{\rm TeV}$ and $171.8~{\rm GeV}\lesssim 
M_t\lesssim173.6~{\rm GeV}$ with the 126 GeV Higgs mass and the correct 
abundance of DM. It must be noted that both the vanishing $\lambda$ and the 
Veltman condition can be satisfied in the region of
 \begin{eqnarray}
  300~{\rm GeV}\lesssim m_S\lesssim1~{\rm TeV},~~
  172~{\rm GeV}\lesssim M_t\lesssim173.6~{\rm GeV}.
  \label{d}
 \end{eqnarray}
We will return to this point later. This DM mass region will also 
be checked by the future XENON100 experiment with 20 times sensitivity of the 
current data~\cite{Cline:2013gha}. One can also see that a larger top mass 
requires a larger DM mass. The reason is similar to the case of the vanishing 
$\lambda$ condition, i.e. a larger top mass needs a larger value of $k$ in order
 to cancel the negative contribution from $-6y^2$ term at the Planck scale, and 
thus a larger $k$ requires a larger DM mass to give the correct abundance in the
 range of $m_S\gtrsim 10^2~{\rm GeV}$.

We also comment on the anomalous dimension for the Higgs mass defined by
 \begin{eqnarray}
  (4\pi)^2\frac{dm_h^2}{dt}=\gamma_{m_h},
 \end{eqnarray}
which indicates the logarithmic divergence. It is also modified to
 \begin{eqnarray}
  \gamma_{m_h}=m_h^2\left(12\lambda+6y^2-\frac{9}{2}g^2-\frac{3}{2}g'{}^2\right)
              +2km_S^2,
  \label{A}
 \end{eqnarray}
where the last term of the right-hand side of Eq.~(\ref{A}) is new contribution 
from the gauge singlet scalar. The value of the anomalous dimension for the 
Higgs mass in the SM at the Planck scale is $(\gamma_{m_h}^{\rm 
SM}/m_h^2)|_{\mu=M_{\rm pl}}\simeq-0.695$. It turns naively out that a singlet mass 
around the EW scale can cancel the negative value of the anomalous dimension 
from the SM. 

In fact, the deep and light blue dots in Fig.~\ref{fig} (a) show a region 
satisfying $|\gamma_{m_h}/m_h^2(M_{\rm pl})|<10^{-2}$ and $\Omega_Sh^2=0.119$ at the 
same time. The deep and light blue dots indicate $10^{-5}<\lambda_S(M_Z)<0.1$ 
and $0.1<\lambda_S(M_Z)<1$, respectively. One can see that 
$|\gamma_{m_h}/m_h^2(M_{\rm pl})|<10^{-2}$ can be satisfied in a region of $200~{\rm
 GeV}\lesssim m_S\lesssim300~{\rm GeV}$ with the corresponding top pole mass, 
$171.8~{\rm GeV}\lesssim M_t\lesssim174.3~{\rm GeV}$. A larger top mass leads a 
smaller value of anomalous dimension due to $12\lambda$ term in Eq.~\eqref{A}. 
Therefore, a larger top mass requires a larger value of $k$ (equivalently to 
$m_S$). However, the magnitude of the decrease of the anomalous dimension by a 
larger top mass is smaller than those of $\lambda$ and Str$M^2/v^2$ because the 
sign of contribution from the top Yukawa coupling only in the anomalous 
dimension is positive unlike the $\lambda$ and Str$M^2/v^2$ cases (see 
Eqs.~\eqref{lam}, \eqref{V}, and \eqref{A}). As a result, a top mass dependence 
of the anomalous dimension is weaker compared to those of the vanishing 
$\lambda$ and the Veltman condition. It should be mentioned that there is also a
 region in which two conditions of the vanishing $\lambda$ and $\gamma_{m_h}$ can
 be realized at the same time.

\subsection{Vanishing beta-function of the self-coupling: {\boldmath $\beta_\lambda(M_{\rm pl})=0$}}

In the SM, the $\beta$-function of $\lambda$ becomes tiny at the Planck scale. 
The value is about $\beta_\lambda^{\rm SM}(M_{\rm pl})\simeq8.42\times10^{-4}$ when 
one takes the Higgs and top pole masses as $m_h=126$ GeV and $M_t=173.07$ GeV at
 low energy. Thus, this condition may also be meaningful for theories beyond the
 SM around the Planck scale. 

For instance, when one takes $m_S=800$ GeV, $k(M_Z)=0.24$, and 
$\lambda_S(M_Z)=0.34$ in addition to $m_h=126$ GeV and $M_t=173.07$ GeV as an 
example in the context of the SM with the gauge singlet field, the corresponding
 values of the $\beta$-function at the Planck scale become $\beta_\lambda(M_{\rm 
pl})\simeq6.32\times10^{-2}$. Therefore, the vanishing $\beta$-function of 
$\lambda$ at the Planck scale in the SM with the gauge singlet cannot be 
satisfied because the runnings of $\lambda$ is increasing from a negative value 
due to the effect of the singlet field as shown in Fig.~\ref{fig} (b). In this 
extension of the SM, $\beta_\lambda(\mu)$ becomes zero at 
$\mu\sim\mathcal{O}(10^{15-17}~{\rm GeV})$ not the Planck scale.

According the above analyses, the BC of $\beta_\lambda(M_{\rm pl})=0$ cannot be 
realized but two BCs of $\lambda(M_{\rm pl})={\rm Str}M^2(M_{\rm pl})=0$ can be 
satisfied in the model. Since the result might indicate that all the Higgs 
potential is induced from a quantum correction under the current circumstances, 
one has no warrant for $\beta_\lambda(M_{\rm pl})=0$. Thus, the non-vanishing 
$\beta$-function can be compatibly understood. Furthermore, there are also two 
additional $\beta$-functions ($\beta_k$ and $\beta_{\lambda_S}$) in this model. 
Since values of $\beta_k(M_{\rm pl})$ and $\beta_{\lambda_S}(M_{\rm pl})$ cannot be 
zero when we impose $\lambda(M_{\rm pl})=0$ and the correct abundance of DM, the 
vanishing condition for only $\beta_\lambda(M_{\rm pl})$ might be meaningless. 
Thus, in this work we take a stance of giving up the vanishing $\beta$-function 
at the Planck scale to predict the Higgs and top masses, although 
$\beta_\lambda(M_{\rm pl})=\lambda(M_{\rm pl})=0$ condition adopted 
in~\cite{Froggatt:1995rt} predicted the values of the Higgs and top masses 
roughly close to experimental magnitudes.

\subsection{Multi coincidence}

It is remarkable that there is a region, given in Eq.~\eqref{d}, satisfying two 
independent BCs at the Planck scale ($\lambda(M_{\rm pl})\simeq0$ and 
Str$M^2(M_{\rm pl})\simeq0$ (or $\gamma_{m_h}(M_{\rm pl})\simeq0$)) at the same time 
with the correct abundance of the gauge singlet DM, 126 GeV Higgs mass, 
experimentally allowed top pole mass, and the coupling perturbativity. This 
double coincidence in the above BCs, $\lambda(M_{\rm pl})\simeq0$ and 
Str$M^2(M_{\rm pl})\simeq0$, with the correct DM abundance is just a non-trivial 
result. The double coincidence means that the Higgs potential becomes flat at 
the Planck scale. The gauge singlet scalar plays a crucial role for the 
realization, and it becomes DM with the correct abundance in the universe at 
present. The double coincidence with DM might be an alternative principle to 
``multiple point criticality principle" discussed in 
Ref.~\cite{Froggatt:1995rt}, where a condition that the SM Higgs potential has 
two degenerate vacua was imposed.\footnote{One vacuum we live is the EW scale, 
and another one is the Planck scale. Under the condition, the vanishing 
$\lambda$ and $\beta_\lambda$ are required.}

\begin{figure}[t]
\hspace{4.5cm}(a)\hspace{8cm}(b)\vspace{-0.7cm}

\begin{center}
\includegraphics[scale=0.98]{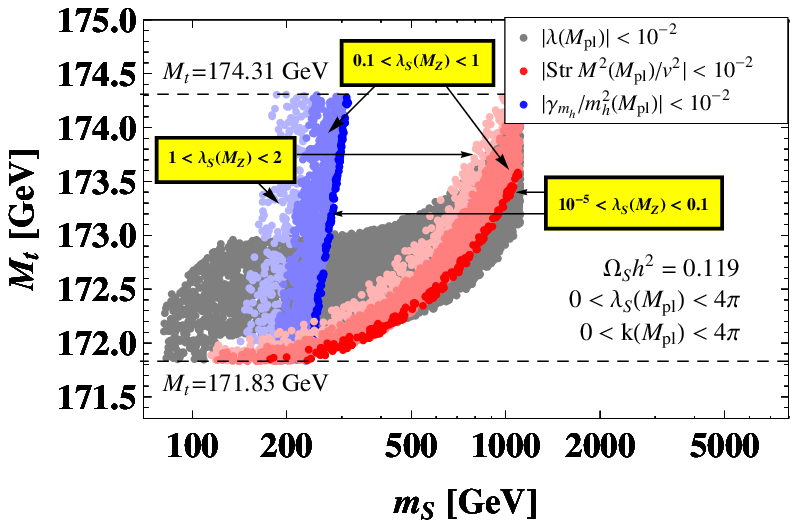}
\includegraphics[scale=0.98,bb=0 0 231 155]{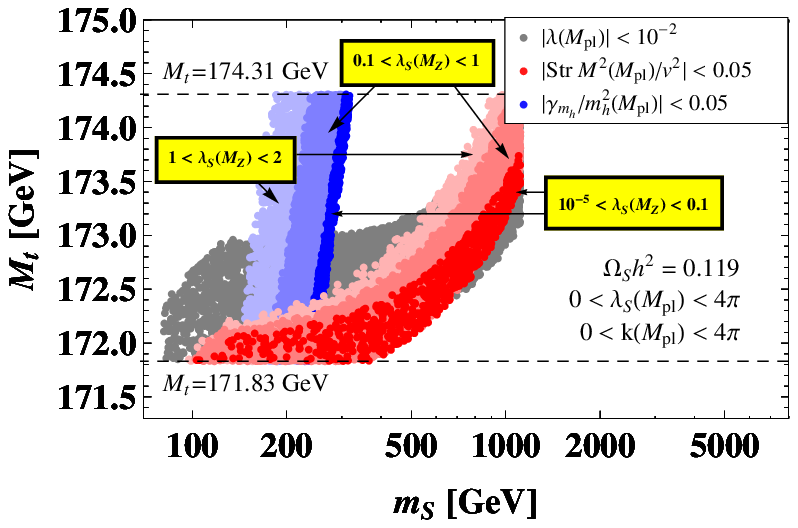}
\end{center}
\caption{(a) Regions satisfying the BCs with $(k(M_{\rm pl}),\lambda_S(M_{\rm 
pl}))<4\pi$, in which conditions $(|\lambda(M_{\rm pl})|,|{\rm Str}M^2(M_{\rm 
pl})/v^2|,|\gamma_{m_h}/m_h^2(M_{\rm pl})|)<10^{-2}$ are depicted as gray, red, and 
blue dots, respectively. The deep, light, and the lightest red (blue) dots 
indicate $10^{-5}<\lambda_S(M_Z)<0.1$, $0.1<\lambda_S(M_Z)<1$, 
$1<\lambda_S(M_Z)<2$ for $|{\rm Str}M^2(M_{\rm pl})/v^2|<10^{-2}$ 
($|\gamma_{m_h}/m_h^2(M_{\rm pl})|<10^{-2}$), respectively. (b) Regions satisfying 
the BCs with $(k(M_{\rm pl}),\lambda_S(M_{\rm pl}))<4\pi$, in which conditions 
$(|\lambda(M_{\rm pl})|,|{\rm Str}M^2(M_{\rm pl})/v^2|,|\gamma_{m_h}/m_h^2(M_{\rm 
pl})|)<(10^{-2},0.05,0.05)$ are depicted as gray, red and blue dots, 
respectively. 
}
\label{fig2}
\end{figure}

In the above analyses, we have limited the values of $k(M_{\rm pl})$ and 
$\lambda_S(M_{\rm pl})$ to be less than 1. But, when one allows the values up to 
$4\pi$, the two regions for the Veltman condition and the vanishing anomalous 
dimension are changed. We also weaken the conditions of $(|{\rm 
Str}M^2(M_{\rm pl})/v^2|,$ $|\gamma_{m_h}/m_h^2(M_{\rm pl})|)<10^{-2}$ to  
$<0.05$, the 
allowed regions for the conditions grow wider. Figure~\ref{fig2} shows the 
cases, and Fig.~\ref{fig2} (a) shows regions satisfying the BCs with $(k(M_{\rm 
pl}),\lambda_S(M_{\rm pl}))<4\pi$ at the Planck scale, in which conditions 
$|\lambda(M_{\rm pl})|<10^{-2}$, $|{\rm Str}M^2(M_{\rm pl})/v^2|<10^{-2}$, and 
$|\gamma_{m_h}/m_h^2(M_{\rm pl})|<10^{-2}$ are depicted as gray, red, and blue dots,
 respectively.
 The deep, light, and the lightest red (blue) dots indicate 
$10^{-5}<\lambda_S(M_Z)<0.1$, $0.1<\lambda_S(M_Z)<1$, $1<\lambda_S(M_Z)<2$ for 
$|{\rm Str}M^2(M_{\rm pl})/v^2|<10^{-2}$ ($|\gamma_{m_h}/m_h^2(M_{\rm pl})|<10^{-2}$), 
respectively. One can see that the region satisfying $|{\rm Str}M^2(M_{\rm 
pl})/v^2|<10^{-2}$ and $|\gamma_{m_h}/m_h^2(M_{\rm pl})|<10^{-2}$ grow wider compared 
to the case shown in Fig.~\ref{fig} (a) when one allows the value of 
$\lambda_S(M_Z)$ up to 2, which corresponds to $\lambda_S(M_{\rm pl})<4\pi$. Such 
a relatively large $\lambda_S(M_Z)$ can effectively increase the value of $k$ 
enough to cancel the negative contribution in the Veltman condition and 
anomalous dimension at the Planck scale even when one takes a smaller $k(M_Z)$. 
In this case, the double coincidence of $\lambda(M_{\rm pl})\simeq0$ and 
Str$M^2(M_{\rm pl})\simeq0$ (or $\gamma_{m_h}(M_{\rm pl})\simeq0$) still occurs.

Figure~\ref{fig2} (b) shows regions satisfying the weaker BCs, $|\lambda(M_{\rm 
pl})|<10^{-2}$, $|{\rm Str}M^2(M_{\rm pl})/v^2|<0.05$, and $|\gamma_{m_h}/m_h^2(M_{\rm
 pl})|<0.05$ with $(k(M_{\rm pl}),\lambda_S(M_{\rm pl}))<4\pi$. In the case, the 
allowed regions become the widest among 
all cases we have investigated. As a
 result, the region satisfying three BCs at the same time appears around 
 \begin{eqnarray}
  150~{\rm GeV}\lesssim m_S\lesssim300~{\rm GeV},~~
  172~{\rm GeV}\lesssim M_t\lesssim172.2~{\rm GeV}.
 \end{eqnarray}
This means that the triple coincidence for the three BCs occurs. The triple 
coincidence requires that the logarithmic divergence of the Higgs mass also 
disappear at the Planck scale instead of allowing a fine-tuning between the bare
 Higgs mass and a quadratic correction.

\section{Summary and discussions}

We have investigated Planck scale BCs on the Higgs sector in the SM with gauge 
singlet scalar DM. The BCs are the vanishing Higgs self-coupling ($\lambda(M_{\rm
 pl})=0$), the Veltman condition (Str$M^2(M_{\rm pl})$=0) (and the vanishing 
anomalous dimension for the Higgs mass parameter, $\gamma_{m_h}(M_{\rm pl})=0$), 
and the vanishing $\beta$-function of the self-coupling ($\beta_\lambda(M_{\rm 
pl})=0$). If one imposes the BCs in the SM, the Higgs and top masses are 
predicted to be close to the experimental ones. BCs of $\lambda(M_{\rm pl})=0$ and
 Str$M^2(M_{\rm pl})$=0 mean that there exists approximately flat direction in the
 Higgs potential. In addition, the quadratic (logarithmic) divergence for the 
Higgs mass disappears under the BC of the Veltman condition (and the vanishing 
anomalous dimension at the Planck scale). However, it actually seems difficult 
to reproduce 126 GeV Higgs mass an $173.07\pm1.24$ GeV top pole mass at the same
 time under the three BCs in the SM. Thus, we have investigated these BCs in the
 context of the SM with the singlet real scalar.

We have taken the setup that the singlet is DM and all scalar quartic coupling 
in the model can be perturbatively treated up to the Planck scale. And we have 
utilized the Higgs with 126 GeV mass in the analyses. We could find that the 
vanishing self-coupling and Veltman condition at the Planck scale can be 
remarkably realized with the 126 GeV Higgs mass and top pole mass, 172 GeV 
$\lesssim M_t\lesssim$ 173.6 GeV and the coupling perturbativity, where a 
correct abundance of scalar dark matter is obtained with mass of 300 GeV 
$\lesssim m_S \lesssim$ 1 TeV. It means that the Higgs potential is 
approximately flat at the Planck scale, and this situation cannot be realized 
in the SM with the top pole mass.

When one takes weaker conditions for the BCs, $(k(M_{\rm pl}),\lambda_S(M_{\rm 
pl}))<4\pi$ and $(|{\rm Str}M^2(M_{\rm pl})/v^2|,$ $|\gamma_{m_h}/m_h^2(M_{\rm 
pl})|)<0.05$, the triple coincidence ($\lambda(M_{\rm pl})\simeq0$, Str$M^2(M_{\rm 
pl})\simeq0$, and $\gamma_{m_h}(M_{\rm pl})\simeq0$) can be realized. 

Next, we discuss some points, which are related with this work and an extension.
 The BC of $\lambda(M_{\rm pl})=0$ implies that our EW vacuum is false and the 
true vacuum is at a high energy scale slightly smaller than the Planck scale 
like the SM with the center values of the Higgs and top masses. We have checked 
that the quantum tunnelling probability $p$ through out the history 
of the universe, which is estimated by $p\simeq V_UH^4{\rm 
exp}(-8\pi^2/(3|\lambda(H)|))$ (e.g., see~\cite{Ellis:2009tp}), can be much 
smaller than 1, where $V_U=\tau_U^4$, $\tau_U$ is the age of the universe as 
$\tau_U\simeq13.7$ Gyrs, and we took $|\lambda(H)|\simeq9.59\times10^{-5}$ with 
$H\simeq8\times10^{17}~{\rm GeV}$ for the true vacuum of our sample point of 
$m_H=126~{\rm GeV}$, $M_t=173~{\rm GeV}$, $m_S=800~{\rm GeV}$, $k(M_Z)=0.24$, and
 $\lambda_S(M_Z)=0.34$.

We comment on the realization of the BCs of $\beta_\lambda(M_{\rm pl})=0$ with 
$\lambda(M_{\rm pl})=0$, which were firstly considered in~\cite{Froggatt:1995rt}, 
in this single extension of the SM. $\beta_\lambda(\mu)$ cannot be zero at the 
Planck scale with $\lambda(M_{\rm pl})=0$ in the extension because there is an 
additional positive contribution from $kS^2|H|^2$ interaction to 
$\beta_\lambda(\mu)$. $\beta_\lambda(\mu)$ becomes zero at 
$\mu\sim\mathcal{O}(10^{15-17}~{\rm GeV})$ (not the Planck scale) with 
$\lambda(M_{\rm pl})\simeq0$ and experimental center values of the Higgs and top 
masses. If one respects both BCs of $\lambda(M_{\rm pl})=\beta_\lambda(M_{\rm pl})=0$
 in this singlet extension of SM, the BCs predict about $145~{\rm GeV}$ Higgs 
mass and $175~{\rm GeV}$ top pole mass at $M_Z$ scale, which are ruled out by 
experiments.
 
Finally, we also comment on other issues such as the existence of the tiny 
neutrino mass and the baryon asymmetry of the universe (BAU), which cannot be 
explained in the SM. One popular explanation is given by adding heavy 
right-handed Majorana neutrinos into the SM. These are known as the seesaw 
mechanism and the leptogenesis for generating the tiny neutrino mass and BAU, 
respectively. In this example of the extension, there exist additional 
contributions from the neutrino Yukawa couplings to $\beta_\lambda$, Str$M^2$, and
 $\gamma_{m_h}$. If the magnitude of the neutrino Yukawa couplings is smaller 
than $\mathcal{O}(10^{-2})$, which corresponds to the right-handed neutrino 
Majorana neutrino mass smaller than $\mathcal{O}(10^{10})~{\rm GeV}$, the 
contributions are negligible in the BCs like the Yukawa couplings of the bottom 
quark and tau. On the other hand, if the neutrino Yukawa couplings are larger 
than $\mathcal{O}(0.1)$, the contributions should be taken account in the BCs. 
For the BC of $\lambda(M_{\rm pl})=0$, a larger $k(M_Z)$ (equivalently a heavier 
DM mass) is required because of a negative contribution from the neutrino Yukawa
 coupling to $\beta_\lambda$. Such a negative contribution may well cancel other 
positive contributions in $\beta_\lambda$ such that $\beta_\lambda(M_{\rm pl})=0$ can
 be realized at the same time. A larger $k(M_Z)$ is needed also for the BC of 
Str$M^2(M_{\rm pl})=0$ because the contribution from the neutrino Yukawa to 
Str$M^2$ is negative. Finally, an effect of the neutrino Yukawa coupling for 
$\gamma_{m_h}$ is relatively non-trivial compared to the other BCs because the 
positive contributions from (top and neutrino) Yukawa couplings to $\gamma_{m_h}$
 compete with the negative one from $12\lambda$ term, i.e. larger (positive) 
Yukawa couplings lead smaller (negative) value of $\lambda$ at the Planck scale.
 Thus, an accurate numerical analysis is required. Effects in the BCs from 
additional particles and their mass scales strongly depend on a model for 
generating the tiny neutrino mass and BAU, e.g. adding right-handed neutrinos, 
but such a model dependent analysis of the BCs with explanations of the neutrino
 mass and BAU in addition to DM might also be interesting.

\subsection*{Acknowledgement}

The authors thank M. Holthausen for answering our question. This work is 
partially supported by Scientific Grant by Ministry of  Education and Science, 
Nos. 00293803, 20244028, 21244036, 23340070, and by the SUHARA Memorial 
Foundation. The works of K.K. and R.T. are supported by Research Fellowships of 
the Japan Society for the Promotion of Science for Young Scientists. The work is
 also supported by World Premier International Research Center Initiative (WPI 
Initiative), MEXT, Japan.

\end{document}